\documentclass[12pt]{article}
  
\usepackage{epsfig}
\usepackage{calrsfs} 
\usepackage{bbm,bm}
\usepackage{cite}
\usepackage{color,cancel,graphicx,mathrsfs}
\usepackage{amsmath}
\usepackage{amssymb}

\usepackage{exscale}

\usepackage{hyperref}
\usepackage{xcolor}
\hypersetup{colorlinks, linkcolor={red!50!black}, citecolor={blue!50!black}, urlcolor={blue!80!black}}

\def\a0size{6}

\newcommand{\ctil}{\widetilde C} 
\newcommand{\ezm}{X _ 1 } 
\newcommand{\intp}{\int \! \! \frac { d ^ 3 p } { ( 2 \pi  ) ^ 3 }} 
\newcommand{\intpp}{\int \!\! \frac { d ^ 3 p' } { ( 2 \pi  ) ^ 3 }} 
\newcommand{\mmeff}{m ^ 2 _ { \rm eff} }

\newcommand{\msub}{m _ {\rm sub}}
\newcommand{\nzm}{X _ 0 } 
\newcommand{\nzmp}{X _ { 0 \perp } }
\newcommand{\phinull}{\varphi} 
\newcommand{\phip}{\hat\Phi} 
\newcommand{\phiq}{\Phi } 
\newcommand{\scalar}{\chi}

\newcommand{\lsi}{\raise0.3ex\hbox{$<$\kern-0.75em\raise-1.1ex\hbox{$\sim$}}}
\newcommand{\gsi}{\raise0.3ex\hbox{$>$\kern-0.75em\raise-1.1ex\hbox{$\sim$}}}

\newcommand{\gsim}{\mathop{\gsi}}
\renewcommand{\vec}[1]{{\bf #1}}

\newcommand{\deltaf} {\,\delta  \! f} 
\newcommand{\ope} {A}
\newcommand{\rhorad} { \rho  }

\begin{document} 

\setlength{\baselineskip}{0.6cm}
\newcommand{\figysize}{16.0cm}
\newcommand{\figtopspace}{\vspace*{-1.5cm}}
\newcommand{\figbottomspace}{\vspace*{-5.0cm}}
  

\begin{centering}
\vfill

{
\centerline{ \Large \bfseries
Scalar field damping at high temperatures
    }
}

\vspace{1cm}

Dietrich B\"odeker\footnote{bodeker@physik.uni-bielefeld.de},
Jan Nienaber\footnote{jan.nienaber@uni-bielefeld.de}

\vspace{.6cm} { \em 
Fakult\"at f\"ur Physik, Universit\"at Bielefeld, 33615 Bielefeld, Germany
}

\vspace{2cm}

\end{centering}

\begin{abstract}
\noindent
The motion of a scalar field that interacts with a hot plasma,
like the inflaton during reheating,
is damped, which is a dissipative process.
At high temperatures the damping can be described by a local term
in the effective equation of motion. 
The damping coefficient is sensitive to multiple scattering. 
In the loop expansion its computation would require an all-order resummation.
Instead we solve an effective Boltzmann equation, similarly to  
the computation of transport coefficients.
For an interaction with another scalar field we obtain 
a simple relation between the damping coefficient
and the bulk viscosity,
so that one can make use of known results for the latter.
The numerical prefactor of the damping coefficient 
turns out to be
rather large, of order $ 10 ^ 4 $.

\end{abstract}
 
\vspace{0.5cm}
\noindent

\vspace{0.5cm}\noindent

 
\vspace{0.3cm}\noindent
 
\vfill \vfill
\noindent
 

 
\section{Introduction}  
\label{s:intro} 

Scalar fields may play an important role in the early Universe.
They can drive cosmic inflation, and their quantum fluctuations 
can provide the seed of galaxy formation, 
they can cause 
phase transitions \cite{Kirzhnits:1972ut} and generate the baryon asymmetry of 
the Universe \cite{Affleck:1984fy}. 
They can also be part or all of dark matter \cite{McDonald:1993ex},
or be responsible for today's dark energy \cite{ Wetterich:1987fm}.

We consider a scalar field $ \varphi  $ which is (approximately)
constant in space and which 
evolves in time. 
Important examples 
are the inflaton field which drives inflation or the axion field
which can be dark matter.
To be specific we will consider $ \varphi  $ being the inflaton,
keeping in mind that our discussion applies to many other situations as well.

When inflation ends,  
the inflaton field $ \varphi  $ starts oscillating
coherently around the minimum of its potential.
It interacts with other fields leading to an energy transfer
thus creating a plasma and (re-)heating the Universe.
At the same time, the motion of $ \varphi  $ gets damped.
The  plasma makes up an increasing fraction of the total energy density.
For sufficiently strong interaction the plasma thermalizes.
Eventually this thermal plasma dominates the energy density; 
the corresponding temperature is called reheat temperature
$ T _ { \rm RH } $.
This is, however,  not the largest temperature of the plasma,
which rises early during the reheating process and then decreases 
before it reaches 
$ T _ { \rm RH } $
\cite{Turner:1983he,Giudice:2000ex}.

An oscillating inflaton field 
with frequency $ \omega  = m _ \varphi  $,
can be viewed as a state with high occupancy of 
inflaton particles with zero momentum and mass $ m _ \varphi  $. 
When there are only few decay products present, 
the damping is dominated by inflaton decay into lighter particles
\cite{Abbott:1982hn}.
If many particles have already been produced such that 
their occupation numbers are of order
one or larger, other effects come into play.
Parametric resonance can lead to very efficient particle production
\cite{Kofman:1994rk}.
Then the  decay products thermalize and acquire a thermal mass.
At high temperature, thermal masses can become larger than
the inflaton mass such that the decay of an inflaton into plasma particles
is  kinematically forbidden~\cite{ Kolb:2003ke}.
Then other processes that  involve multiple
scatterings become unsuppressed and open new channels for the energy 
transfer~\cite{Drewes:2013iaa,Anisimov:2010gy}.
In this paper we consider 
the damping rate in the 
high-temperature regime 
where $ T $ is much larger than $ m _ \varphi  $ and the mass of the
plasma particles.
We assume that the characteristic frequency 
$ \omega  \sim \dot \varphi  /\varphi  $  and the damping rate $ \gamma  $
of $ \varphi  $ 
are small compared to the thermalization rate 
of the plasma.
Then the inflaton interacts nearly adiabatically with an almost thermal 
plasma. 
In particular, 
there is no non-perturbative particle production through parametric resonance
\cite{Kofman:1994rk}.
When the plasma is approximately thermal, its properties are fully specified
by the temperature and by the instantaneous value of $ \varphi  $.
Therefore the plasma `forgets' about its past, and its effect
on the inflaton dynamics can be described by local terms.
The effective equation of motion (without Hubble expansion) 
for the 
zero-momentum mode of $ \varphi  $ takes the form 
\cite{Hosoya:1983ke} 
\begin{align}
	\ddot \varphi  + V' _ { \rm eff } 
	+ \gamma 
	\dot \varphi  = 0
	\label{effeom} 
	,
\end{align} 
where the prime denotes a derivative with respect to $ \varphi  $.
The effective potential $ V _ { \rm eff } $ and the damping coefficient 
$ \gamma  $ only depend on the value of $ \varphi  $ and the temperature.
For sufficiently slow evolution, higher derivative terms 
in Eq.~(\ref{effeom}) can be neglected. 
Note that the form of Eq.~(\ref{effeom}) follows from
the separation of timescales alone.

If $ \omega  $ and  the damping rate $  \gamma  $ are small compared 
to the thermalization rate, 
then $ \gamma  $
can be obtained from a finite-temperature real-time correlation function, 
evaluated in the the zero-momentum, zero-frequency limit
\cite{McLerran:1990de,Bodeker:2006ij}.
E.g., the damping coefficient for the axion field is proportional
to the Chern-Simons diffusion rate in QCD \cite{ McLerran:1990de}, 
the so-called strong sphaleron rate, which is non-perturbative
and has been calculated on the lattice \cite{ Moore:1997im,Moore:2010jd}. 
In many cases the required  correlation functions
can in principle be calculated perturbatively in thermal field theory.
They are, however, sensitive to
timescales much larger than  the mean free time
of the plasma particles, so that one 
has to take into account multiple scatterings.
This requires the resummation of an infinite set 
of diagrams. 
Several authors have applied 1- or 2-loop approximations with 
resummed propagators containing a finite width 
(see e.g. \cite{Morikawa:1984dz,Morikawa:1985mf,Gleiser:1993ea,Berera:1998gx}), 
which gives rise to a nonzero damping rate.
However, proper treatment of the multiple interaction requires the resummation
of a much larger class of diagrams \cite{Jeon:1992kk,Bastero-Gil:2010dgy}.

Similar complications arise in the computation of transport coefficients, 
such as bulk viscosity,
which can be written as the zero-momentum zero-frequency limit of
a stress-tensor correlation 
function~\cite{Jeon:1994if,jeon:1995zm}.
For viscosities, the required summation of diagrams has been 
performed.
It was shown that this is equivalent  
to solving an effective Boltzmann equation \cite{Jeon:1994if}. 

The  physics behind bulk viscosity $ \zeta  $ 
is closely related to the one of 
the damping coefficient $ \gamma  $.
Both describe 
small deviations from thermal equilibrium.
In the case of $ \zeta  $ it is  
due to a uniform expansion of the system.
The deviation of the trace of the stress tensor 
from the ideal-fluid form is proportional to $ \zeta  $.
When the field $ \varphi  $ changes with time and interacts with 
the plasma particles, it changes their parameters such as masses 
or couplings, driving the plasma out of equilibrium.
Since $ \varphi  $ is spatially constant, this deviation
from equilibrium is homogeneous and isotropic as well.

For the model considered in Ref.~\cite{ Bodeker:2006ij} 
the damping coefficient could be related to a 
correlation function of the stress tensor. 
Thus there is a simple relation between $ \gamma  $ and the  
bulk viscosity $ \zeta  $ of the thermal plasma, 
so that one can use the known result for $ \zeta  $.
In this work we consider  interactions of $ \varphi  $ with another scalar 
field $ \chi  $ through operators 
that cannot be expressed in terms the stress tensor, but which 
still  allow for a perturbative treatment.
We can proceed similarly to the computation of the bulk viscosity,
for which the required
resummation of diagrams is equivalent to solving an 
effective Boltzmann equation \cite{Jeon:1994if}. 
Damping coefficients have been computed from a Boltzmann equation
long ago using a relaxation time approximation 
\cite{Hosoya:1983ke}.%
\footnote{ The damping coefficients computed in  
Refs.~\cite{Morikawa:1985mf,Gleiser:1993ea,Berera:1998gx} are of
the same form as in Ref.~\cite{Hosoya:1983ke}.
} 
This approximation, however, does not give the correct result for the 
bulk viscosity in scalar theory
\cite{Jeon:1994if,jeon:1995zm}.
Here we carefully treat the collision term as well as thermal effects
by employing the effective Boltzmann equations which were used to
perturbatively compute bulk viscosities
in scalar theories
\cite{Jeon:1994if,jeon:1995zm},
and in gauge theories\cite{Arnold:2006fz,Laine:2014hba}.
This allows us to obtain the correct dependence on the coupling constants
and explicitly compute $ \gamma  $ at leading order in perturbation theory.

This paper is organized as follows. 
In Sec.~\ref{s:eom} we obtain the effective equation of
motion for $ \varphi  $ and an expression for the damping 
coefficient in terms of the plasma-particle occupancy. 
The latter is computed in Sec.~\ref{s:bol}	from an effective 
Boltzmann equation.
In Sec.~\ref{s:dam} the solution to the Boltzmann equation is inserted
into the effective equation of motion for $ \varphi  $,
and the damping coefficients is expressed in terms of
the known bulk viscosity of the plasma.
Section \ref{s:con} contains conclusions and a brief outlook.
Appendix  \ref{a:dTdm}  deals with the 
thermodynamics of the plasma particles, and 
Appendix  \ref{a:sol} describes the solution of the Boltzmann equation.

\section{Effective equation of motion}
\label{s:eom}

In this section, largely following Ref.~\cite{Hosoya:1983ke}, 
we obtain the effective equation of motion (\ref{effeom}) 
from quantum field theory and relate the coefficients therein to 
microscopic physics.
We consider a scalar 
field $ \Phi  $ is coupled to another scalar field $ \scalar  $ 
through the interaction
\begin{align}
	{ \cal L } _ { \phiq \scalar}  = - \ope ( \phiq)   \scalar  ^ 2 
	\label{int} 
	.
\end{align} 
Restricting ourselves to renormalizable interactions 
we can have
\begin{align} 
	\ope ( \phiq) = \frac \mu  2  \phiq+ \frac \lambda  4 \phiq^ 2 
	\label{ope} 
\end{align} 
with coupling constants $ \mu $ and $ \lambda  $.
Without Hubble expansion 
the equation of motion for $ \phiq$ reads
\begin{align}
	\ddot \phiq
	- \Delta  \phiq
	+ V'( \phiq) 
	+ \ope'( \phiq) \scalar ^ 2
	= 0
	\label{eom} 
	,
\end{align}
where $ V $ is the part of the tree-level potential that depends only on
$ \phiq$.
Eq.~(\ref{eom}) is still an equation for field operators.
We want to write an equation of motion for the zero-momentum mode 
$ \varphi  $ of $ \Phi  $,
and we assume that $ \varphi  $ can be approximated by a classical field. 
We write 
\begin{align} 
	\phiq = \phinull + \phip
	\label{decomp}
\end{align} 
where $ \phip $ contains the non-zero momentum modes of $ \phiq $.
Through the interaction, $ \chi  $ particles are produced. 
Once the $ \chi  $ particles are created, they can also produce 
$ \phiq $ particles which are represented by $ \phip $.
Thus the production of $ \phiq $ particles also contributes to the damping
of $ \phinull $.
This effect is discussed in 
\cite{Ema:2016kpf}, where it was found that this contribution is 
subdominant unless the energy density in $ \phinull $ is small compared to 
the energy density of  the $ \chi  $ particles.
In the context of reheating after inflation this would already be 
during radiation domination.
We assume that $ \phinull $ still dominates the energy 
density and neglect this contribution.
Then we can replace the forth term in Eq.~(\ref{eom}) by $ A'( \varphi  )
\chi  ^ 2 $.

We assume that $ \scalar  $ interacts rapidly with itself or other fields,
so that it thermalizes on timescales which are short compared
to the period of $ \phinull $ oscillations.
Furthermore, we assume that the interactions of $ \scalar $ 
are weak enough, so that
the typical mean free path of $ \scalar $ particles 
is much larger than their typical de Broglie wavelength.
Then $ \scalar  $ is made up of  weakly interacting 
particles
which can be described by their phase space density, or occupancy 
$ f ( t, \vec p )  $.
Since we consider a homogeneous system, it only depends on time $ t $ and
on the particle momentum $ \vec p $.
We may then replace $ \scalar ^ 2 $ by its expectation value computed
from the occupancy using the 
free-field expression
\begin{align}
	\left \langle \scalar  ^ 2 \right \rangle 
	=
	\intp \frac {  f (t, \vec p ) } E 
 	\label{fluc}
	,
\end{align} 
where $ E $ is the one-particle energy (see below).
Thus we arrive at the effective {\it classical}  equation of motion
\begin{align} 
	\ddot \phinull   
	+ V'( \varphi  ) 
	+ \ope'( \varphi )   \langle \scalar  ^ 2 \rangle 
	= 0
	.
	\label{eomeff}
\end{align}

The deviations from equilibrium are assumed to be small, so that
the occupancy in Eq.~(\ref{fluc}) can be written as
\begin{align} 
	f ( t, \vec p ) = 
	f _ { \rm eq } ( t, \vec p ) 
	+ \deltaf ( t, \vec p ) 
	\label{fapp} 
\end{align} 
with the local equilibrium distribution 
\begin{align}
	f _ { \rm eq } ( t, \vec p ) =
	\frac 1 { \exp \big ( E/T   ) - 1 }
	\label{feq} 
	.
\end{align} 
and $ \delta  f \ll f _ { \rm eq } $.
The temperature $ T $ in Eq.~(\ref{feq}) varies slowly with time. 
The mass of the $ \scalar $ particles depends on the value of $ \varphi  $,
\begin{align}
	m ^ 2 = m _ 0 ^ 2 + 2 \ope (   \varphi ) 
	\label{m2} 
\end{align}
where  $ m _ 0 $ is the zero-temperature mass at vanishing $ \varphi  $.
Throughout this paper we assume that $ m $ is small compared
to the temperature.%
\footnote{
	The opposite limit 
	$ m \gg T $ is considered in Ref.~\cite{ Mukaida:2012qn} 
	with additional light degrees of freedom. 
	Then $ \scalar $ can be integrated out giving rise to an effective
	interaction of $ \varphi  $ with the light plasma particles.
} 
The mass appearing in the 
one-particle energy $ E $  
in Eqs.~(\ref{fluc}) and~(\ref{feq}) 
also receives a thermal contribution $ m ^ 2  _ { \rm th } \propto
T ^ 2 $, so that 
$ E = ( \vec p ^ 2 + m ^ 2 _ { \rm eff } ) ^{ 1/2 } $ with
\begin{eqnarray}
    m ^ 2 _ { \rm eff } = m ^ 2 +   m ^ 2 _ { \rm th } 
   .
   \label{meff2} 
\end{eqnarray} 
To avoid a tachyonic instability \cite{Drees:2021wgd},
$  m ^ 2 _ { \rm eff } $ must be positive.

With the help of 
Eq.~(\ref{fapp}), 
the expectation value in Eq.~(\ref{fluc}) becomes
\begin{align}
	\langle \scalar ^ 2 \rangle = \langle \scalar ^ 2 \rangle _ { \rm eq }
	+ \delta  \langle \scalar ^ 2 \rangle 
	\label{dchi} 
	.
\end{align}
The first term in Eq.~(\ref{dchi}) is nondissipative. 
It gives a thermal correction in the effective potential 
in Eq.~(\ref{effeom}) \cite{Hosoya:1983ke},
\begin{align}
	V ' _ { \rm eff } &= V ' + A' \langle \chi  ^ 2 \rangle _ { \rm eq } 
	\label{veff} 
	,
\end{align} 
which is precisely the leading term in the high-temperature limit 
of the 1-loop effective potential
(see, e.g., \cite{Laine:2016hma}).
The second term in 
Eq.~(\ref{dchi}) is dissipative and will give rise to the damping term
in Eq.~(\ref{effeom}). 

\section{Boltzmann equation }
\label{s:bol} 

The occupancy of $ \chi $ particles in Eq.~(\ref{fluc}) can be computed
by solving a Boltzmann equation, because the timescale on which 
their mass changes is of order $ 1 / \omega  $ which is much
larger than their typical de Broglie wavelength of order $ 1/T $. 
Due to the homogeneity, spatial momentum is conserved.
Thus the Boltzmann equation takes the form
\begin{align}
	\label{bol}
	\partial _ t f = C 
\end{align} 
where $ C $ is the collision term.
Now we insert Eq.~(\ref{fapp})
on the left-hand side of Eq.~(\ref{bol}).
We neglect $  \partial _ t   \deltaf $
because it is quadratic in small quantities, so that 
\begin{align}
	\partial _ t f \simeq 
	 - f _ { \rm eq } ( 1 + f _ { \rm eq } ) 
	\partial _ t  ( E/T )  .
	\label{fdot}
\end{align} 
The zero-momentum mode $ \phinull $ depends on time and changes the 
mass of the plasma particles through the interaction \eqref{int}. 
If the oscillation is much slower than the thermalization of the
plasma, this is an adiabatic process
that changes the temperature  in Eq.~(\ref{feq}) at constant volume.%
\footnote{
	This does not apply to the case $ \omega  \gsim  
	T $ which is considered in 
	Refs.~\cite{ Drewes:2014pfa,Laine:2021ego,Ming:2021eut}.
}
Thus the time dependence of the temperature is determined by
\begin{align} 
	\partial _ t T
	=
	\left ( \frac { \partial T } { \partial m ^ 2 } \right )
	_ {\!\!\! S, V }  
	\partial _ t m ^ 2 .
\end{align} 
In the limit $ T \gg m $ we obtain
(see Appendix \ref{a:dTdm}) 
\begin{align}
	\left ( \frac { \partial T } { \partial m ^ 2 } \right )
	_ {\!\!\! S, V }  
	=
	\frac T { 4 \rhorad    } 
	\left \langle \scalar ^  2 \right \rangle _ {\!\rm eq } 
     \label{dTdm}
     ,
\end{align} 
where $ \rhorad  $ is the energy density of the thermal plasma.
The one-particle energy $ E $ depends on time through the effective mass.
We thus have
\begin{align}
	\partial _ t ( E/T ) 
	= 
	\frac { 1  } { 2 T E } 
	\left [   
		1 
     -\frac { \left \langle \scalar ^  2 \right \rangle _ {\!\rm eq } } 
								{ 2 \rhorad  } 
		\left ( 
			 E  ^ 2 - T ^ 2 
				\frac { \partial \mmeff }
					 { \partial T ^ 2 }
          \right )
	\right ]  
	\partial _ t m ^ 2  .
	\label{inner} 
\end{align} 
The third term in the square bracket is small compared to the first, both
for hard ($ | \vec p | \sim T $)
and for soft ($ | \vec p | \sim m _ { \rm eff }  $)
momenta, and can be neglected,
so that 
\begin{align} 
	\partial _ t f 
	\simeq 
      - f _ { \rm eq } ( 1 + f _ { \rm eq } ) 
	\frac Q { 2 T } 
	\partial _ t m ^ 2 
	\label{fdotfi}
\end{align} 
with 
\begin{align} 
	Q ( \vec p ) 
	 \equiv  
	\frac 1 E - \frac { \langle \scalar ^ 2 \rangle _ { \rm \! eq} }
	{ 2 \rhorad  } E.
	\label{S}
\end{align}

Now we insert Eq.~(\ref{fapp}) into the collision term.
Since $ C $ vanishes in equilibrium, its expansion  in $ \delta  f $
starts at linear order, 
\begin{align} 
	C \simeq \widehat C \deltaf 
	.
	\label{clin} 
\end{align} 
Here we have neglected  the contribution of $ \phiq $ particles 
because the corresponding collision term is quadratic in the 
$ \Phi $-$ \chi $ couplings which we assume to be much smaller than
the self-coupling of $ \chi $ entering $ \widehat C $.

It is convenient to write the deviation from equilibrium as 
\begin{align} 
	\deltaf = - f _ { \rm eq } \left ( 1 + f _ { \rm eq }  \right )
	X
	\label{deltaf}
	.
\end{align} 
Similarly we write the  linearized collision term as
\begin{align}
	  \widehat C \deltaf  
	= 
	f _ { \rm eq }  (  1 + f _ { \rm eq } ) 
	 	\widetilde  C X  
	\label{ctil}
	,
\end{align} 
with the convolution 
\begin{align} 
	[ \widetilde  C X ] ( \vec p ) 
	\equiv
	\intpp
	\widetilde C ( \vec p, \vec p ')  
	f _ { \rm eq } ( \vec p') \Big (  1 + f _ { \rm eq } ( \vec p' ) \Big )  
	X ( \vec p') 
	.
\end{align} 
Then the  kernel $ \widetilde C $ is symmetric \cite{Arnold:2003zc},
\begin{align} 
	 \widetilde C ( \vec p , \vec p') 
	= \widetilde C ( \vec p', \vec p ) 
	.
	\label{sym}
\end{align} 
The Boltzmann equation thus turns into an equation for $ X $, 
\begin{align}
	- \frac {   \partial _ t m ^ 2} { 2 T } Q=	
	\widetilde C X 
	\label{bolp} 
	.
\end{align} 
Since the collision term vanishes in equilibrium for any temperature,
the linearized collision term has a zero mode 
$ X = \ezm $ 
associated with 
a shift of the temperature, which is given by $ \ezm ( \vec p )  = E $. 
Due to the symmetry of $ \ctil $ 
the right-hand side of Eq.~(\ref{bolp}) 
is orthogonal to $ \ezm $.
For Eq.~(\ref{bolp})  to be consistent,  the left-hand side 
must be orthogonal to $ \ezm $ as well. 
This is indeed 
the case when the second term in Eq.~(\ref{S}) is taken into account,
which can be easily checked.

Due to the zero mode the linear operator $ \widetilde C $ cannot be
inverted.
However, it can be inverted 
on the subspace orthogonal to the zero mode,%
\footnote{%
This is equivalent to imposing the  Landau-Lifshitz condition 
on the energy density
$ \delta \rhorad = 
	 ( 2 \pi  ) ^ {-3 } 
  \int d ^ 3 p E \deltaf 
= 0 $.
}
where orthogonality is defined with respect to the inner product
\begin{align}
	( X   , X  ' ) 
	\equiv
	\intp  f _ { \rm eq } ( 1 + f _ { \rm eq }  ) 
	X   ( \vec p )
	X  ' ( \vec p ) 
	\label{prod} 
	.
\end{align} 
We can then write the solution as%
\begin{align}
	X  = - \frac { \partial _ t m ^ 2}  {2 T}  	
	\widetilde C ^{ -1 }  Q 
	\label{chisol}
	,
\end{align}    
and we finally obtain
\begin{align}
\delta f=f_{\rm eq}(1 + f_{\rm eq})\frac{\partial_t m^2 }{2T} 
	\widetilde{C}^{-1}Q
	.
	\label{dfs} 
\end{align}

\section{Damping coefficient and bulk viscosity}
\label{s:dam} 

Coming back to the effective equation of motion (\ref{eomeff}), we
insert the solution (\ref{dfs}) into Eq.~(\ref{fluc}) to
obtain the second term in Eq.~(\ref{dchi})	 as
\begin{align} 
	\delta  \langle \scalar ^ 2 \rangle 
	=	
	\frac 1 { 2 T } \left ( Q' , \widetilde C ^{ -1 } Q \right ) 
	\partial _ t m ^ 2 
	\label{ds2p}
     .
\end{align} 
Here we have introduced 
$ Q'( \vec p ) \equiv 1/E $.
The factor $ \partial _ t m ^ 2 $ is proportional to $ \dot \varphi  $.
Comparing Eqs.~(\ref{effeom}) and  (\ref{eomeff}) we see that 
the second term in Eq.~(\ref{dchi}) is indeed responsible for the damping,
\begin{align} 
	\gamma  \dot \varphi  &= A' \delta  \langle \chi  ^ 2 \rangle
	\label{grel} 
	.	
\end{align} 
Inserting $ A $ from Eq.~(\ref{ope}) we obtain
\begin{align}
	\gamma  = \frac 1 { 4 T } 
	\left ( Q' , \widetilde C ^{ -1 } Q \right ) 
	\left ( \mu  + \lambda  \varphi  \right ) ^ 2
	\label{g} 
\end{align} 
which is our main result.

The computation of $ \widetilde C ^{ -1 } Q $ is described
in Appendix  \ref{a:sol}. 
However, at this point we do not need it explicitly,%
\footnote{ It will be usefull later, when we estimate the size of 
$ \deltaf $ in order to check the accuracy of our approximations.}
because, as we shall see in a
moment,  the 
coefficient $ ( Q' , \widetilde C ^{ -1 } Q ) $ 
also appears in the computation of the bulk viscosity~$ \zeta  $ of the
$ \chi $ plasma.
Therefore it can be read off directly from known results for $ \zeta  $. 
To see this, we first recall that 
$ \widetilde C ^{ -1 } Q $ is orthogonal to $ \ezm = E$, i.e., 
$ ( E, \widetilde C ^{ -1 } Q ) = 0 $.
In Eq.~(\ref{g}) we may therefore replace 
$ Q'= Q + (  \langle \scalar ^ 2 \rangle _ { \rm \! eq} /
{ 2 \rhorad  } ) E $ by $  Q $ without changing our result for $ \gamma  $,
which then reads
\begin{align}
	\gamma  = \frac 1 { 4 T } 
	\left ( Q , \widetilde C ^{ -1 } Q \right ) 
	\left ( \mu  + \lambda  \varphi  \right ) ^ 2
	\label{gQ} 
     .
\end{align}  

Let us now recall some properties of the bulk viscosity, as described, e.g., 
in Ref.~\cite{Arnold:2006fz}. 
When a plasma is uniformly 
compressed or rarified
it leaves equilibrium, unless this happens infinitely slowly. 
The pressure of the plasma then differs from the value it would
have in the equilibrium state with the  same energy density. 
This deviation of the pressure from equilibrium is proportional to the bulk 
viscosity.

In a plasma with scale invariance the bulk viscosity vanishes, for two
different reasons.
The first one is that a uniform expansion or rarefaction is a dilatation
which is a symmetry transformation in a scale invariant theory.
Therefore such a transformation does not take the system out of equilibrium.
The second is that in a 
scale invariant theory the trace of the energy-momentum 
tensor $ T ^{ \mu  \nu  } $ always vanishes.
Therefore the pressure $ P = T ^{ m m } / 3 $ equals $ \rho  /3 $ even
out of equilibrium.

Scale invariance is broken by zero-temperature masses and by the
trace anomaly, i.e., by quantum effects.
The bulk viscosity is then quadratic in the measure which 
controls  the breaking of scale invariance. 

The bulk viscosity $ \zeta  $ of the thermal plasma of scalar particles 
with mass $ m $
was computed for scalar theory in 
Refs.~\cite{Jeon:1994if,jeon:1995zm}.
Like in QCD \cite{Arnold:2006fz} it can be written as
\begin{align}
	\zeta  = \frac 1T
	\left ( q , \widetilde C ^{ -1 } q \right  ) 
	\label{zet}
	, 
\end{align}
with 
\begin{align}
	q ( \vec p ) =
     -\frac 1E
	\left [ \left ( c _ { \rm s } ^ 2 - \frac 13 \right ) 
		 \vec p ^ 2 + c _ { \rm s } ^ 2 \msub ^ 2  
	\right ] 
	\label{q}
	.
\end{align} 
Here $ c _ { \rm s } $ with $ c _ { \rm s } ^ 2 = \partial P/\partial \rho  $
is the speed of sound.
In a scale invariant theory $ c _ { \rm s } ^ 2 $ equals $   1/3 $, so that the
first term in the square bracket in Eq.~(\ref{q}) vanishes.
Furthermore, $ \msub $ with
\begin{align}
   \msub ^ 2 \equiv \mmeff 
   - T ^ 2 \frac { \partial \mmeff  } { \partial T ^ 2 }
\end{align}
is the so-called subtracted mass.
In the massless limit $ m = 0 $,  $ \mmeff$ equals $ T ^ 2 $ times 
a function of the coupling constants (see Eqs.~(\ref{m2}), (\ref{meff2})).
Then the only contribution to the subtracted mass is from the running
of the couplings renormalized at the scale $ T $.
The subtracted mass is thus a measure of the deviation from scale 
invariance as well, 
because it vanishes  when $ m =0 $ and the couplings do not run.
Since $ q $ appears twice in Eq.~(\ref{zet}), the bulk viscosity
is indeed quadratic in the measure of scale-invariance violation.

Now we replace 
$ \vec p ^ 2 $ by $  E ^ 2 - \mmeff $ in Eq.~(\ref{q}) which turns it into
\begin{align}
	q ( \vec p ) =   
	\left [ \left ( c _ { \rm s } ^ 2 - \frac 13 \right ) \mmeff
		 - c _ { \rm s } ^ 2 \msub ^ 2  
	\right ] 
     \frac 1E
	- \left ( c _ { \rm s } ^ 2 - \frac 13 \right ) E
   \label{qE}
	.
\end{align} 
Comparing Eqs.~(\ref{S}) and \eqref{qE} we see that 
both $Q $ and $ q $  consist of a term 
proportional to $ 1/E $, and one proportional to $ E $. 
Furthermore, $ q $ appears 
on the left-hand side of a Boltzmann equation precisely like
$ Q $ in Eq.~(\ref{fdotfi}),%
\footnote{See Eq.~(3.7) of Ref.~\cite{Arnold:2006fz}.
} 
and is thus 
orthogonal to $ \ezm $ as well. 
Therefore $ q $ must be proportional to $ Q $. 
Here we are interested in the limit  $ T \gg m $
in which \cite{Jeon:1994if}
\begin{align}  
   |  c _ { \rm s } ^ 2 - 1/3 |  =   O ( m ^ 2 _ { \rm sub  }/T ^ 2) 
   \ll 1   
   \qquad (T\gg m ) 
   , 
\end{align} 
and also $ \mmeff \ll T ^ 2 $. 
Therefore we can approximate the square bracket in (\ref{qE})
by~$- \msub ^ 2/3 $.
This gives us the approximate factor of proportionality,
so that $ q \simeq -(  \msub ^ 2 /3 )  Q$. 
Then we obtain the following simple relation 
\begin{align} 
	\gamma  ( \varphi  , T) 
	=
	\frac  9  4 \frac \zeta  { \msub ^{ 4 } }	
	( \mu  + \lambda   \varphi  ) ^ 2
	\label{ga} 
\end{align} 
of the damping coefficient in the effective equation of motion 
(\ref{effeom}) and the bulk viscosity of the $ \chi $ plasma.
Like in Ref.~\cite{Bodeker:2006ij} the nontrivial dependence 
on the interaction of the plasma particles, 
on thermal masses, etc., is precisely the same for both quantities. 
Note that $ \msub ^ 4 $ in the denominator of Eq.~(\ref{ga}) 
removes the factors related to the breaking of scale invariance 
from $ \zeta  $, which can also be seen explicitly in Eqs.~(\ref{ze}) and
(\ref{ega}) below.
Thus, despite its similarity to the bulk viscosity, the damping
coefficient is not related to the breaking of scale invariance.

The bulk viscosity for a self-interacting scalar field
was computed in Ref.~\cite{Jeon:1994if}. 
For the quartic self-interaction
\begin{align}
	{ \cal L } _ { \scalar\scalar} = - \frac { g ^ 2 }
	{ 4! } \scalar^ 4
	\label{self} 
\end{align} 
and $ T \gg m  $ 
the leading-order result reads 
\begin{align}
	  \zeta  = \frac b 4 
	\frac {\msub ^ 4\mmeff  } { g ^ 8 T ^ 3 } 
	\ln ^ 2 
	\left ( \frac { \kappa ^ 2 \mmeff  } { T ^ 2 } \right )
	\label{ze} 
\end{align}  
with $ b = 5.5\times 10 ^ 4 $ and $ \kappa  =1.25 $. 
The effective mass for the $\scalar$ particles is given by 
\begin{align}
	m_{\rm eff}^2= m ^ 2 + \frac{g^2}{24}T^2
	, \qquad 
	m^2=m_0^2+\frac{\lambda}{2}\phinull^2 
	.
	\label{meff} 
\end{align}
Inserting Eq.~(\ref{ze}) into Eq.~(\ref{ga}) we obtain the damping coefficient
\begin{align}
	\gamma ( \varphi, T  )   
	= a 
	\frac {\mmeff  } { g ^ 8 T ^ 3 } 
	\ln ^ 2 
	\left ( \frac { \kappa ^ 2 \mmeff  } { T ^ 2 } \right )
	( \mu  + \lambda  \varphi  ) ^ 2
	\label{ega} 
\end{align} 
with the remarkably large numerical prefactor 
\begin{align}
	a = 3.1\times 10 ^ 4 
	\label{a} 
	.
\end{align}
In the temperature range $ T \gg m /g ^ 2 $ the form (\ref{ze}) and 
thus Eq.~(\ref{ega}) remain valid when a
cubic self-interaction is included in (\ref{self}),
while in the intermediate regime 
$ m \ll T \ll m/g ^ 2 $ the bulk viscosity depends nontrivially
on the relative strengths of cubic and quartic $ \chi  $ self-couplings
\cite{ Jeon:1994if}.
It is obvious from the dependence on the coupling constant $ g $ that 
the result \eqref{ega} cannot be
obtained from a one-loop approximation to a  correlation function, 
as anticipated  
in Refs.~\cite{Bastero-Gil:2010dgy,Mukaida:2012qn}.   
Instead, by solving the Boltzmann equation we have summed an infinite
set of diagrams which all contribute at leading order in $ g $. 

We can compare our result with the one obtained in Ref.~\cite{Hosoya:1983ke}
for a  single scalar field 
by putting $ \scalar = \varphi  $, $ \mu  = 0 $, and,
up to an $ O ( 1 ) $
factor, $ g ^ 2 = \lambda $.  
In Ref.~\cite{Hosoya:1983ke}  the Boltzmann equation was solved
in the collision-time approximation, i.e., by replacing
the linearized collision term on the right-hand side of
Eq.~(\ref{clin}) by a constant times $ \delta  f $.
Such an approximation does not take into account the zero eigenvalues
and the hierarchy of nonzero eigenvalues of $ \widetilde C$.
In Ref.~\cite{Hosoya:1983ke}  the collision time
is determined by $ 2 \to 2 $ scattering which changes momenta
but not particle numbers.
Bulk viscosity and the damping coefficient $ \gamma  $ are, 
however, determined by the slowest equilibration
process, corresponding to the smallest eigenvalue of the
linear collision operator,
since it is the inverse of the collision operator that appears in
Eqs.~(\ref{g}) and  \eqref{zet}. 
In scalar field theory the slowest process is particle number 
equilibration.
Therefore the computation of Ref.~\cite{Hosoya:1983ke} does not 
give the correct dependence on the coupling constant and 
underestimates the values of $ \gamma  $ and $ \zeta  $. 
Similarly, in Ref.~\cite{Yokoyama:1998ju} the rate for 
elastic scattering was used to estimate the damping coefficient.

The importance of particle number changing processes for the bulk viscosity
is well known. 
The reason why they are also important for the damping coefficient is the 
following. 
When $ \varphi  $ evolves in time, 
it changes the mass of the $ \chi $~particles,
but not their momenta.
This, in turn, changes the energy density of $ \chi $~particles
but leaves their number density unaffected. 
In order to relax to equilibrium, the $ \chi $ particle number has to adjust
to the equilibrium value corresponding to their new energy density. 

Let as finally discuss  the range of validity of 
the effective equation of motion \eqref{effeom}. 
There the damping term is linear in $ \dot \varphi   $. 
This is related to the linearization of the Boltzmann equation, which
requires that $ \delta  f \ll f _ { \rm eq } $. 
In Appendix  \ref{a:sol} we show that  
\begin{align}
	 \delta  f / f _ { \rm eq } 
	\sim
	\frac { m _ { \rm eff } } { g ^ 8 T ^ 4 } \partial _ t m ^ 2
	\label{dfof} 
\end{align} 
for the interaction (\ref{self}).
When $ \varphi  $ oscillates with frequency $ \omega  $ and amplitude
$ \widetilde \varphi  $, 
the time derivative $ \partial _ t m ^ 2 $ can be estimated as 
$ \lambda   \omega  \widetilde \varphi  ^ 2 $.
For $ g T \gg m $ we thus need
\begin{align} 
	\lambda  \omega   \widetilde \varphi  ^ 2 \ll g ^ 7 T ^ 3
	\label{pcon} 
\end{align}
in order to be able to linearize the Boltzmann equation.

We may also apply the condition \eqref{pcon} 
to a model with a single scalar field  
$ \varphi  $ which was considered in Ref.~\cite{Hosoya:1983ke} 
by putting $ \mu  = 0 $ and  $ \lambda  \sim g ^ 2 $.
Then \eqref{pcon} turns into $ \omega  \widetilde  \varphi  ^ 2
\ll g ^ 5 T ^ 3 $. 
The energy density in  $ \varphi  $ would be 
$ \rho  _ \varphi  \sim \omega  ^ 2 \widetilde \varphi ^ 2
\ll  (\omega  /T) g ^ 5 T ^ 4 $.
Due to $ \omega  \ll T $, the energy carried by $ \varphi  $  would be 
only a tiny fraction of the plasma energy density $ \rho  \sim T ^ 4 $.
For the more interesting case that we have several fields,
$ \varphi  $ can give the 
dominant contribution to the total energy without violating 
the condition~\eqref{pcon}.

\section{Conclusion } 
\label{s:con} 

A slowly moving
homogeneous scalar field $ \varphi  $ interacting with a thermal plasma
drives it slightly out of equilibrium, 
giving rise to dissipation and damping. 
In the high-temperature regime
the damping coefficient in the effective equation of motion for $ \varphi  $  
can be efficiently computed by solving an appropriate Boltzmann
equation, see Eq.~(\ref{g}).
We have considered a plasma made of a single species of scalar particles.
In this case we obtained a simple relation of the damping coefficient
to the bulk viscosity of the plasma, Eq.~(\ref{ga}). 
This extends a result  
\cite{Bodeker:2006ij}
which was obtained for a scalar field with derivative interaction.
Like in the computation of viscosity, the solution of the Boltzmann
equation is dominated by the slowest process required for equilibration.
This can be easily generalized to multicomponent plasmas, where
again one has to identify the slowest process to solve the Boltzmann 
equation and then use the resulting phase space density to 
compute the dissipative terms in the effective equation of motion
for the scalar field.

\section*{Acknowledgements}
We thank Mikko Laine and Simona Procacci for comments and discussions. 
D.B. acknowledges support by the Deutsche Forschungsgemeinschaft 
(DFG, German Research Foundation) through the CRC-TR 211 
'Strong-interaction matter under extreme conditions'– project number 
315477589 – TRR 211. 

 \appendix
\renewcommand{\theequation}{\thesection.\arabic{equation}}

\global\long\def\theequation{\thesection.\arabic{equation}}

\section{%
Mass dependence of the temperature
}
\label{a:dTdm}
\setcounter{equation}0
The free energy of an ideal gas has the high-temperature expansion
\begin{align}
	F ( T, V, m ^ 2 ) 
 	= 
	V( -a T ^ 4 + b T ^ 2 m ^ 2 + \cdots  ) 
\end{align} 
with positive constants $ a $ and $ b $; 
$ m $ is the mass of one particle species. 
The coefficient $ a $ can also contain the contributions from 
other light species.
At leading order our expansion is related to the energy density by
$ \rho  = 3 a T ^ 4 $.
For a scalar
\begin{align}
	\frac { \partial F } { \partial m ^ 2 }
	=
	\frac V2 \langle \scalar ^ 2 \rangle _ { \rm eq }  
	\label{dFdm}
\end{align} 
which gives
$ b = \langle \scalar ^ 2 \rangle _ { \rm eq }  /( 2 T ^ 2 ) $. 
The entropy is
\begin{align}
	S = - \frac { \partial F } { \partial T } 
	= V ( 4 a T ^ 3 - 2 b T m ^ 2 + \cdots  ) .
	\label{entro} 
\end{align}
This can be inverted to obtain the expansion for the temperature,
$ T = T _ 0 + T _ 2 + \cdots  $, for which we obtain
\begin{align} 
	T _ 0 &= \left ( \frac S { 4 a V } \right ) ^{ 1/3 } ,
    \label{T0} 
	\\
	T _ 2 &= \frac b { 6a } \frac { m ^ 2 } { T _ 0} 
   \label{T2} 
   .
\end{align} 
Differentiating $ T _ 2 $ with respect to $ m ^ 2 $ then
gives Eq.~(\ref{dTdm}).

\section{%
Solving the Boltzmann equation and estimating~$ \delta  f $  
}
\label{a:sol}
\setcounter{equation}0


The linearization of the Boltzmann equation is only possible 
if the deviation from equilibrium is small, $\delta f \ll f _ { \rm eq } $.
This condition  restricts
the allowed values of the couplings  and the amplitude of the
zero-momentum mode $ \varphi  $.

To estimate the size of $ \delta  f $ in Eq.~(\ref{dfs})  
we derive its explicit form, closely following Ref.~\cite{jeon:1995zm}.
For a self-interacting scalar field 
one has to include two contributions
in the collision term, 
\begin{align} 
	\ctil = \ctil _ { \rm el }  + \ctil _ { \rm inel } 
	\label{csum}
	.
\end{align}  
$ \ctil _ { \rm el }  $ 
describes elastic $ 2 \to 2 $ scattering which conserves particle number.
Therefore it has the  additional 
zero mode $ \nzm = 1  $, associated with a shift of the chemical potential,
and 
cannot be inverted on the subspace orthogonal to $ \ezm = E $. 
One also  has to include an 
inelastic contribution   $ \ctil _ { \rm inel }  $
describing particle number changing processes, even though its matrix
element is higher order.
$ \ctil $ has a single small eigenvalue $ c $ on the subspace 
orthogonal to $ \ezm $,
with the approximate eigenvector 
\begin{align}
	 \nzmp = \nzm - \alpha  \ezm
	\label{nzmp} 
\end{align} 
where $ \alpha  = ( \ezm, \nzm ) /( \ezm, \ezm ) $.
The small eigenvalue is approximately
\begin{align}
   c = \frac { \big ( \nzmp, \ctil _ { \rm inel } \nzmp \big) } 
       { ( \nzmp, \nzmp) } 
   \label{c} 
\end{align} 
while the other nonvanishing eigenvalues are of order $ \ctil _ { \rm el } $.
In the numerator of Eq.~(\ref{c})	we may replace $ \nzmp $ by $ X _ 0 $
because $ \ctil _ { \rm inel } X _ 1 $ vanishes.
This eigenvalue gives the leading contribution to $ \ctil ^{ -1 } $, so that
\begin{align}
	 \ctil ^{ -1 } Q  
	\simeq
	\frac { ( \nzmp , Q  ) }
		{ ( \nzm , \ctil _ { \rm inel }  \nzm ) }
     \nzmp 
	.
	\label{ciq} 
\end{align} 
In the numerator of Eq.~(\ref{ciq})  we can replace $ \nzmp $ by $ \nzm $,
because $ X _ 1 $ is orthogonal to $ Q $. 
We insert this into Eq.~(\ref{dfs}), which finally gives 
\begin{align}
\deltaf ( \vec p ) 
   =f_{\rm eq}( \vec p ) 
   \big[1 + f_{\rm eq}( \vec p )\big ]  \frac{\partial_t m^2 }{2T} 
	\frac { ( \nzm , Q  ) }
		{ ( \nzm , \ctil _ { \rm inel }  \nzm ) }
     \nzmp ( \vec p ) 
	.
	\label{dff} 
\end{align}
We now use this to estimate the size of $ \delta  f $.
We will encounter the integrals
\begin{align}
	I _ n \equiv \intp  f _ { \rm eq } ( 1 + f _ { \rm eq }  ) E ^ n
   \label{In} 
\end{align}   
for $ n = -1, 0 $, and $ 1 $. For $ n \ge 0 $ these are saturated
at $ | \vec p |  \sim T $, giving $ I _ n \sim  T ^ { 3+n } $.
Since $ f _ { \rm eq } \simeq T/E $ for $ E\ll T $, 
the integral $ I _ { -1 } $,  is logarithmically infrared 
divergent in the massless limit and is cut off by~$ m _ { \rm eff } $. 
Thus $ I _ { -1 } $  receives leading order contributions both 
from $ | \vec p | \sim T $
and from $ | \vec p |  \sim m _ { \rm eff } \ll T $, with the result
\begin{align} 
   I _ { -1 } = \frac { T ^ 2 } { 2 \pi  ^ 2 } 
   \ln \left ( \frac { 2 T }  { m _ { \rm eff } }  \right )
   \label{Im1} 
   .
\end{align}  
The factor $ ( \nzm , Q  ) $ 
in the numerator of Eq.~(\ref{dff}) contains $ I _  { -1 } $ and $ I _ 1 $
and is of order $ T ^ 2 $ modulo logarithms, because
$ \langle \chi  ^ 2\rangle _ { \rm eq } / \rho  \sim T ^ { -2 } $. 
The denominator depends on the type of interaction (see below).

Since the size of $ \deltaf ( \vec p ) $ depends on $ | \vec p | $,
we need to know which values of $ | \vec p | $ give the dominant contributions
to $ ( Q' , \widetilde C ^{ -1 } Q ) \propto ( Q', \nzmp ) $ which enters the
damping coefficient in Eq.~(\ref{g}). 
Using Eq.~(\ref{nzmp}) we find that the integrals \eqref{In}  
appear in the combination $ I _ { -1 } - \alpha  I _ 0 $.
The factor $ \alpha  $ is of order $ 1/T $.
Thus  $ | \vec p | \sim  T $ and 
$  | \vec p | \sim m _ { \rm eff } $ are equally important.
In both regions $  \nzmp \sim 1 $.
Due to the Bose factors in Eq.~(\ref{deltaf})  the ratio 
$ \delta  f / f _ { \rm eq } $ increases with decreasing $ | \vec p | $, 
so that it takes its largest value when $ | \vec p | \sim m _ { \rm eff } $.
Putting $ | \vec p | \sim m _ { \rm eff } $, 
collecting all factors and ignoring logarithms we obtain 
\begin{align}
\deltaf /f _ { \rm eq } 
   \sim  
    \frac { T ^ 2 \partial _ t m ^ 2 } 
          { m _ { \rm eff } ( \nzm , \ctil _ { \rm inel }  \nzm ) } 
	\label{ddf} 
   .
\end{align}

For the $ \chi  $ self-interaction (\ref{self}) 
and the $ \scalar $-$ \varphi  $ interaction \eqref{int} with $ \mu  =0 $,
$\ctil _ { \rm inel } $ describes scattering involving 6 particles.
The corresponding squared matrix is proportional to~$ g ^ 8 $.
The momentum integral which enters the 
denominator in Eq.~(\ref{ciq}) is saturated by soft momenta 
($ T \sim  m _ { \rm eff })$)   \cite{Jeon:1994if}.
It contains up to six Bose distributions, which for soft momenta satisfy 
$ f _ { \rm eq } ( \vec p ) \simeq T/E $, giving rise to a factor $ T ^ 6 $. 
By dimensional analysis one then finds
$ ( \nzm , \ctil _ { \rm inel }  \nzm ) 
\sim g ^ 8 T ^ 6/m _ { \rm eff } ^ 2 $,
which yields the estimate \eqref{dfof}.

\bibliography{references}
\bibliographystyle{jhep}

\end{document}